\documentstyle[epsf,aps]{revtex}
\draft
\begin{document}
\title{Control of Decoherence and Relaxation by Frequency Modulation of Heat Bath}
\author{G.S. Agarwal \footnote{also at Jawaharlal Nehru Centre for Advanced
Scientific Research, Bangalore, India.}}
\address{Physical Research Laboratory, Navrangpura, Ahmedabad-380 009, India.}
\date{\today}
\maketitle
\begin{abstract}
We demonstrate in a very general fashion, considerable slowing down of
decoherence and relaxation by fast frequency modulation of the system heat bath coupling. The
slowing occurs as the decoherence rates are now determined by the spectral
components of bath correlations which are shifted due to fast modulation.  We
present several examples including the slowing down of the heating of a trapped
ion, where the system - bath interaction is not necessarily Markovian.
\end{abstract}
In recent times the decoherence of a coherent superposition state has acquired a
new dimension \cite{{one},{two},{three},{four},{five}} because of the requirement of
the stability of such a superposition. The stability has been investigated for
certain systems. The decoherence rates have been calculated and even measured
\cite{six} in the context of Cat like states \cite{seven} for the radiation
field in a cavity. The decoherence issues are also very significant in the context of quantum
computation \cite{{eight},{nine},{ten}}. Clearly the stability of coherent
superpositions requires methods for slowing down the decoherence. Several
proposals exist in the literature \cite{{eleven},{twelve},{thirteen}}. These involve
for example use of a sequence of pulses\cite{twelve} or engineering of the density
of states associated with the reservoir or even changing the reservoir
interaction from a single photon to multiphoton (or more generally multiboson)
interaction \cite{{eleven},{fourteen}}. Other proposals
involve feedback methods \cite{thirteen}.
It may be added that spontaneous emission in many systems is also a cause of 
decoherence. We now understand reasonably well, how to inhibit spontaneous 
emission either by manipulating the density of states \cite{fifteen} or by using external 
fields \cite{{sixteen},{seventeen}}. The methods based on external
fields could be especially useful for slowing down decoherence.

In this letter, we discuss a method based on the frequency modulation
\cite{{nineteen},{twenty},{twenty one}} of the system-heat bath (environment) coupling. We specifically
assume large frequency modulation and take the modulation index $m$
to have a value given by $J_0 (m) = 0$. Under these conditions, we demonstrate
considerable slowing down of the decay and decoherence rates. We present a physical basis 
for this slowing down. We present several examples including the heating of a
trapped ion. Our method is useful only if the correlation time $\kappa^{-1}$
of the heat bath is larger than the rate of frequency modulation i.e. 
$\nu^{-1}$. It must be noted that a recent experimental proposal to control 
decoherence [6(b)] also depends on a coherent coupling with a bath (second
single mode cavity) and works under similar conditions \cite{note}.

In order to appreciate the basic idea of using frequency modulation we consider
a two state system,  where the states $|a\rangle$ and $|b\rangle$ are coupled by some field. We also assume that the
state $|b\rangle$ decays at the rate $2\kappa$. This simple model [Fig.1] can describe many
physical situations. For example, it can represent an excited atom in a cavity
\cite{twenty two} in which the photon leaks out at the rate $2\kappa.$ In this case the states
$|a\rangle$ and $|b\rangle$ will correspond to $|e,o\rangle$, $|g,1\rangle,$ where
$|o\rangle$ and $|1\rangle$ represent vacuum and one photon state respectively and where
$|e\rangle$ and $|g\rangle$ represent the excited and ground states of the atom.  It can also describe a situation where the state $|b\rangle$ could be an 
excited state coupled to ionization continuum.  The
probability amplitudes $C_a$ and $C_b$ for the states $|a\rangle$ and
$|b\rangle$ obey the equations
\begin{eqnarray}
\dot{C}_a&=& - igC_b,\nonumber\\
\dot{C}_b&=& - \kappa C_b - ig^*C_a.
\end{eqnarray}
We have removed any fast time dependence by working in an appropriate frame. If 
$\kappa$ is large, then as is well known
\begin{eqnarray}
|C_a|^2 &\cong& \exp\{-2\Gamma t\}, \nonumber \\
\Gamma &=& |g|^2/\kappa ;~ \kappa\gg  g.
\end{eqnarray}
The decay of the state $|a\rangle$ arises from the decay of the state
$|b\rangle$. In the opposite limit $(\kappa\rightarrow 0)$ one gets oscillatory behavior, which in the cavity context
is known as the vacuum field Rabi oscillation. We now consider the effect of a
phase modulation $m\sin\nu t$ on the decay of the state $|a\rangle:$ we assume a
modulation of the coupling constant 
\begin{equation}
g\rightarrow g\exp\{- im\sin\nu t\}.
\end{equation}
Here $m$ and $\nu$ give respectively, the amplitude and the frequency of the
modulation.  Equation (1) is no longer amenable to analytical solutions. In
Fig. 2,  we display the excited state population for different values of $\nu$
and $m$ chosen to be a zero of the Bessel function of order zero 
\begin{equation}
J_o (m) = 0.
\end{equation}
This choice of $m$ will become clear in the analysis to follow. In Fig. 2 
we also show the behavior in the absence of modulation. We observe that under the
condition(4) the decay of the excited state population is considerably slowed as
the modulation frequency increases. This clearly demonstrates {\it how a frequency
modulation can slow down the effects of decay.} We thus have a {\it method of
controlling relaxation / decay} by frequency modulation.
We now explain the observed numerical behavior for large $\nu$. Using (1), we
can easily derive the following integro-differential equation for the amplitude
of the excited state
\begin{equation}
\dot{C_a}\equiv -|g|^2 e^{-i\Phi(t)}\int_0^t e^{-\kappa(t-\tau)+i\Phi(\tau)}C_a
(\tau)d\tau.
\end{equation}
We use
\begin{equation}
e^{-i\Phi(t)} = \sum_{l=-\infty}^{+\infty} J_l(m)e^{-il\nu t},
\end{equation}
and we assume that (i)~ $\nu$ is large (ii)~$ C_a (\tau)$ varies slowly with~
$\tau$~and carry out a long time average denoted by over bar to get
\begin{eqnarray}
\frac{\partial}{\partial t}{\rm(lnC_a)}\equiv -|g|^2 \int_0^t e^{-\kappa\tau}
\overline{e^{-i\Phi(t)}e^{i\Phi(t-\tau)}}~ d\tau,\nonumber \\
\equiv -|g|^2\sum_{p}J_p^2 (m)(\kappa+ip\nu)^{-1}.
\end{eqnarray}
In order to slow down the decoherence, we need to remove the $\nu=0$ term in
(7). This can be achieved by imposing the condition (4)  whence (7) reduces to  
\begin{equation}
\frac {\partial}{\partial t}{\rm(ln C_{a})}
\approx -2|g|^2 J_1^2(m)(\kappa^2 +\nu^2)^{-1}\kappa.
\end{equation}
Therefore the decay of the excited state
occurs at a modified rate $2\tilde\Gamma$ with
\begin{equation}
\frac{\tilde\Gamma}{\Gamma}\cong J_1^2
(m)\left(\frac{2\kappa^2}{\kappa^2+\nu^2}\right).
\end{equation}
The decay factor (9) agrees very well with the behavior shown in the Fig. 2 for
 $20\pi$ as then $\nu \gg \kappa$. The very fast oscillations do not show up on the scale
of the Fig. 2. The result (9) can be understood by noting that - (i) the factor
in the parenthesis in (9) is just the factor that one would have obtained with
a detuned interaction between the states $|a\rangle$ and $|b\rangle$; (ii)
the Bessel function represents the strength of the first side band.

We next demonstrate that the above idea applies rather generally. We consider
the usual microscopic treatment of the heat bath \cite{twenty
three}
with the {\it modulation of the system heat bath coupling}. For the purpose of
illustration, we consider a spin system [raising and lowering operators $S^+$ and
$S^-$] interacting say with a dc and ac magnetic field in Z direction so that
the unperturbed Hamiltonian is $(\omega_0 - m\nu \cos \nu t)S^z$. The energy
separation gets modulated - such modulations are routinely used (see e.g.
Noel {\it et al} \cite{nineteen}). In the interaction picture the interaction 
with the heat bath can be written as 
\begin{equation}
H_I(t)=(S^+e^{+i\omega_0 t-i\Phi(t)} R^-(t)+{\rm H.c.}),
\end{equation}
where $R^-(t)$ is the appropriate operator for the heat bath. As usual
\cite{twenty three} we will assume that the coupling of the bath to the 
system is weak. The heat bath is characterized in terms of the correlation 
functions:
\begin{eqnarray}
\langle R^-(t)\rangle &=& 0,\nonumber\\
\langle R^+(t+\tau) R^-(t)\rangle &=& C^{+-} (\tau),\nonumber\\
\langle R^-(t+\tau) R^+(t)\rangle &=& C^{-+} (\tau),\nonumber\\
\langle R^-(t+\tau) R^-(t)\rangle &=& 0.
\end{eqnarray}
The Fourier transforms of~$C^{+-}$~ and~$ C^{-+}$~ are related via the fluctuation
dissipation theorem. We can now do the standard calculation \cite{twenty three} to derive a master
equation for the reduced density matrix $\rho$ of the system alone. We quote the
result of this calculation.
\begin{eqnarray}
\frac{\partial \rho}{\partial t} = -(S^+S^-~\rho -S^-~\rho S^+)
\int_0^t d\tau ~C^{-+}(\tau)e^{+i\omega_0\tau}e^{-i\Phi(t)}e^{i\Phi
(t-\tau)},\nonumber\\
-(\rho~S^-S^+ - S^+~\rho~ S^-)\int_0^t d\tau~C^{+-} (-\tau)e^{+i\omega_0\tau}
e^{i\Phi(t-\tau)-i\Phi(t)},\nonumber\\
{\rm +~terms~with~subscripts}~\pm~\rightarrow~\mp, ~\omega_0 ~\rightarrow
-~\omega_0,~\Phi~\rightarrow -\Phi.
\end{eqnarray}
First of all we note, that if the bath correlations were like delta
correlations ~$C^{-+}(\tau) = 2\delta (\tau)C^{-+}$,~ then the master equation (12)
does {\it not depend on the modulation~$\Phi$}. Clearly, the {\it bath correlation
time $\tau_c$ has to be at least of the order of the time} associated with the
modulation. Under the fundamental condition (4), the time average in (12) can be 
{\it approximated by}
\begin{equation}
\overline{e^{-i\Phi(t)+i\Phi(t-\tau)}} \cong 2J_1^2 (m)~\cos~\nu~\tau
\end{equation}
and then (12) reduces to 
\begin{eqnarray}
\frac{\partial\rho}{\partial t}\equiv -2(S^+S^- \rho - S^-\rho~ S^+)
\int_0^\infty d\tau C^{-+}(\tau)e^{+ i\omega_0\tau}~\cos~\nu~\tau~ J_1^2
(m),\nonumber\\
-2~(~\rho~ S^-S^+ - S^+ \rho~S^-)\int_0^\infty d\tau C^{+-}
(-\tau)e^{+i\omega_0\tau}~\cos~\nu~\tau~ J_1^2(m),\nonumber\\
{\rm +~terms~ with}~\pm~\rightarrow~\mp,~\omega_0~\rightarrow~-\omega_0.
\end{eqnarray}
The standard master equation corresponds to the limits $\nu\rightarrow 0, J_1^2
\rightarrow 1$. It is clear that if $\nu$ is large enough compared to frequency
scale of $C^{-+}(\tau)e^{i\omega_0\tau}$ then the real part of the integral in
(14) will be approximately zero and {\it decoherence effectively does not
exist}. In particular, if~ $C^{-+}(\tau) ~= ~C_0^{-+}~ e^{-\kappa\tau-i\omega\tau},
~C^{+-}(-\tau)~=~ C_0^{-+}~e^{-\kappa\tau - i\omega\tau},$~ then 
\begin{eqnarray}
\frac{\partial\rho}{\partial t} &=& -\frac{2(\kappa-i\Delta)J_1^2(m)}{(\kappa -
i\Delta)^2+\nu^2} \{C_0^{-+}(S^+S^-~\rho - S^-~\rho~ S^+) \nonumber \\
&+& C_0^{+-}(\rho~S^-S^+ - S^+\rho~S^-)\}{\rm + c.c.}.,\Delta = (\omega_0 -
\omega).
\end{eqnarray}
Clearly, the relaxation coefficients in the master equation are modified by
factors like (9). Hence the relaxation is much slower. In particular, the
relaxation of the coherence~$\langle S^\pm\rangle$~ will be on a much longer time
scale. For large $\nu$ compared to $\kappa$ and $\Delta$, the relaxation time is very
large.

We next consider the application of the above ideas to the decoherence of an ion
in a trap. This is important in many applications of ion traps such as in
connection with the production of Cat states and in quantum computation. In
particular, we consider the possibility of {\it reducing} the {\it heating} of the {\it ground state of trapped
ions}. In a recent letter, James \cite{ten} has considered a model for heating
produced by a stochastic field {\rm E(t)}. In terms of the annihilation and 
creation operators $a, a^{\dagger}$ associated with the ionic motion the heating
is described by the Hamiltonian
\begin{equation}
H_1 = i\hbar[u(t)a^\dagger - u^*(t)a],
\end{equation}
where $u(t) = iZE(t)e^{i\omega_0 t}/\sqrt{2M\hbar\omega_0}$.  The field $E(t)$ is
a Gaussian stochastic process. The time scale of the stochastic field is taken
to be comparable to the time scale of the ionic motion.   Hence this model is
{\it outside} the usual {\it markovian limit}. We now consider the effect of an
external modulation so that effectively,
$u(t)\rightarrow u(t)e^{-i\Phi(t)}$. Following James' work the fidelity 
${\rm F(t)}$ of the ground state is given by
\begin{equation}
F(t)=[1+2\langle|v(t)|^2\rangle + \langle|v(t)|^2\rangle^2 - |\langle
v^2(t)\rangle|^2]^{-1/2},
\end{equation}
\begin{equation}
v(t) = \frac{iZ}{\sqrt{2M\hbar\omega_0}}\int_0^t E(t^\prime)e^{-i\Phi(t^\prime)+i\omega_0
t^\prime}dt^\prime.
\end{equation}
The mean values in (17) can be obtained from (18) by assuming exponential
correlation for $E(t):$
\begin{eqnarray}
\langle u(t)u^*(t^\prime)\rangle &=&\frac{\Omega^2}{2}
e^{-\kappa|t-t^{\prime}|}\nonumber\\
\langle |v|^2 \rangle &=&\frac{\Omega^2}{2}\sum_{- \infty}^{+\infty}\sum J_n (m) 
J_p (m) I (\omega_0-n\nu ~; ~-\omega_0+p\nu) \nonumber \\
\langle v^2\rangle &=& - \frac{\Omega^2}{2}\sum_{-\infty}^{+\infty}\sum J_n (m) J_p 
(m)I(\omega_0 - n\nu ~;~ \omega_0 - p\nu) 
\end{eqnarray}
where the integral $I(\omega_\alpha, \omega_\beta)$ is found to be
\begin{eqnarray}
I(\omega_\alpha,\omega_\beta) &=& (i\omega_\alpha + i\omega_\beta)^{-1} 
\left[(\kappa + i\omega_\beta)^{-1}e^{i(\omega_\alpha+\omega_\beta)t}
-(\kappa - i\omega_\alpha)^{-1} \right]\nonumber \\
&+& (i\omega_\alpha - \kappa)^{-1}(-i\omega_\beta -\kappa)^{-1}
e ^{i\omega_\alpha t- \kappa t}\nonumber \\
&+& ~{\rm terms~ with}~ \alpha\Leftrightarrow\beta.
\end{eqnarray}
Note that $\omega_\alpha + \omega_\beta$ can vanish in which case, a limiting procedure leads
to
\begin{equation}
I (\omega_\alpha, -\omega_\alpha) =  
(\kappa - i\omega_\alpha)^{-1}~ t + (i\omega_\alpha - \kappa)^{-2}
\left(e^{i\omega_\alpha t-\kappa t}-1\right) + {\rm c.c.}
\end{equation}
We show the fidelity factor $F$ in Fig. 3, both in the absence and presence of
the modulation. We choose a parameter domain in which {\it fidelity}
was being {\it degraded} rather fast. Clearly, if we assume large frequency
modulation and condition (4), then as the figure shows, there is considerable
{\it improvement in the fidelity} under frequency modulation of the stochastic field
{\rm E(t)} responsible for heating the trapped ion.

In conclusion, we have shown how the appropriate modulation of the system-heat
bath interaction can slow down the decay as well as the decoherence to a very 
large extent.  This happens 
as generally, the decoherence is determined by the spectral components of
the bath correlation functions. If system-bath interaction is modulated, then
the decoherence is determined by the spectral components, which are shifted by
the multiples of the modulation frequency. If the modulation frequency is large
compared to the width of the bath correlations, then we would get much smaller
decoherence rate. Finally, note that we have a method to control the effects of 
decoherence since the modulation depth and frequency can be
varied.

The author thanks Sunish Menon for the plots.


\begin{figure}
\caption{Schematic representation of a generic two level system with lower level decay
into a bath. A strong modulation of the coupling slows the decay of the upper
level.}
\end{figure}
\begin{figure}
\caption{The Population in the excited state $|C_a(t)|^2$ (Eqs.(1),(3)) as a
function of $t$ for different values of the modulation frequency and for
$\kappa=10g$. The modulation index is chosen to be the fifth zero of (4). The
curves from top to bottom are for $\frac{\nu}{\pi g} = 20,5,0.5$ and $0.05$. The
curve for $m=0$ is hardly distinguishable from the curve for $\frac{\nu}{\pi
g}=0.05$.}
\end{figure}
\begin{figure}
\caption{Fidelity (17) or the heating of an ion in the trap for two
different values of the modulation frequency $\nu = 5, 3$. The dashed curve
gives the result in the absence of the modulation. Parameters are chosen as
$\omega_0/\kappa =1 ;~\Omega$ [defined by Eq.(19)]$ =\sqrt2\omega_0$. The wiggles
arise from the periodic modulation.}
\end{figure}
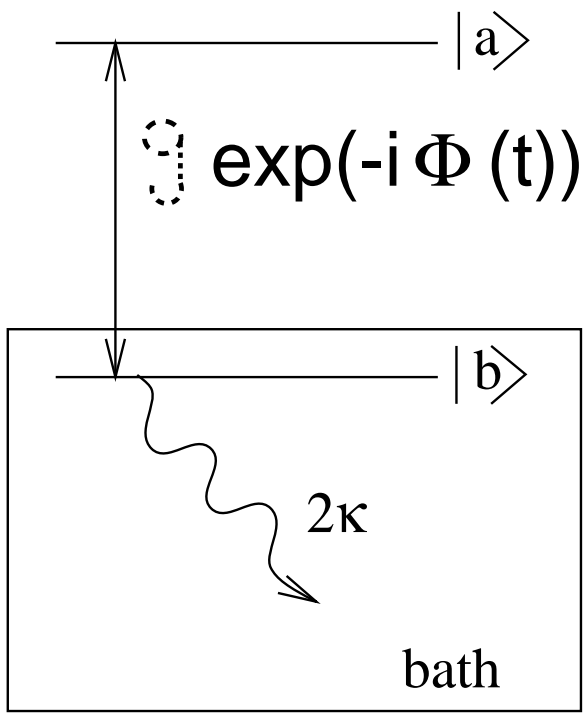
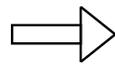
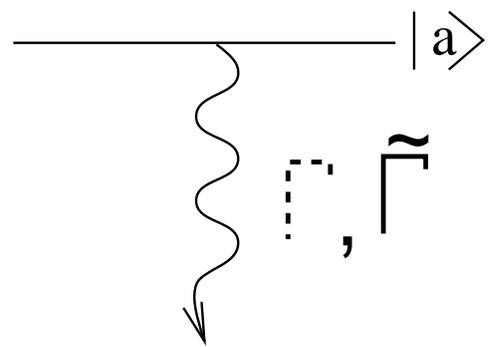
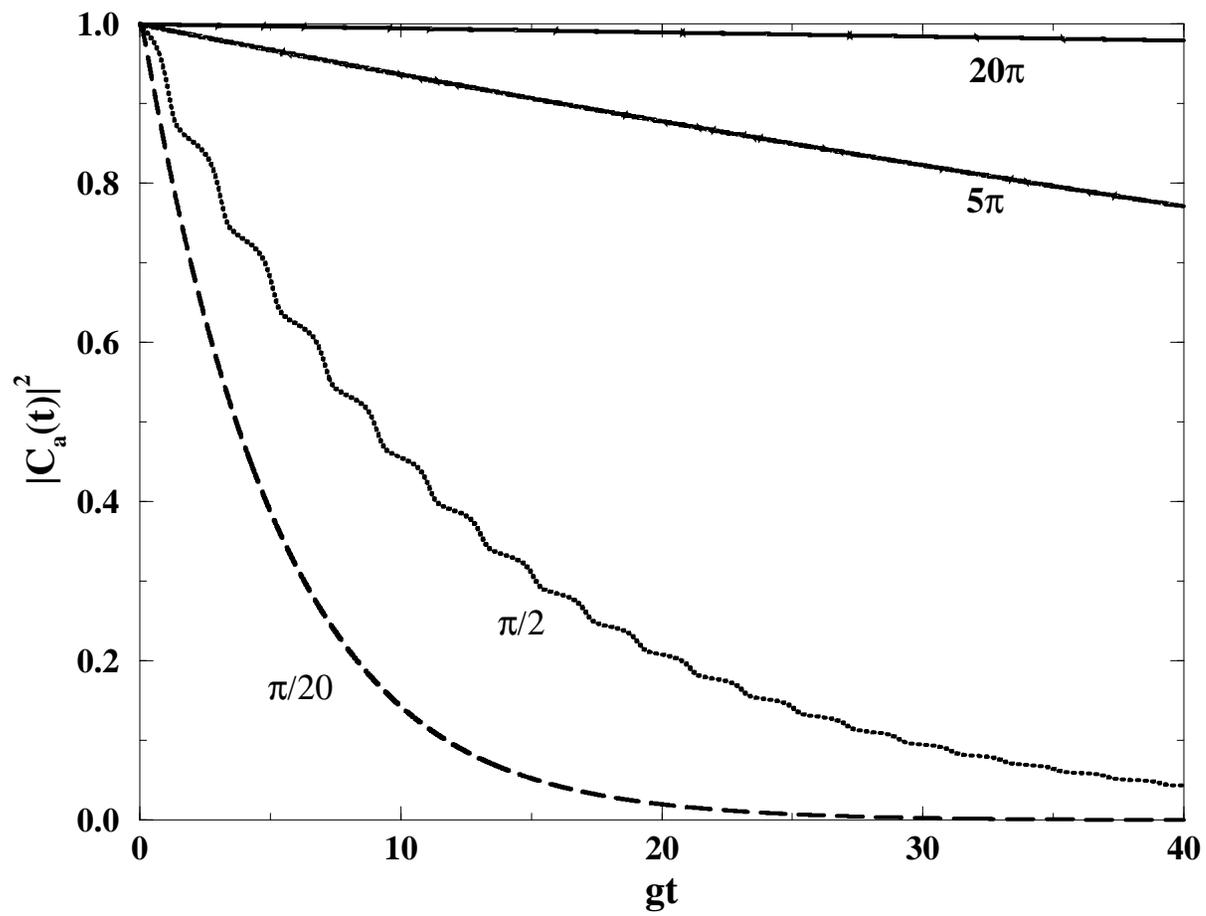
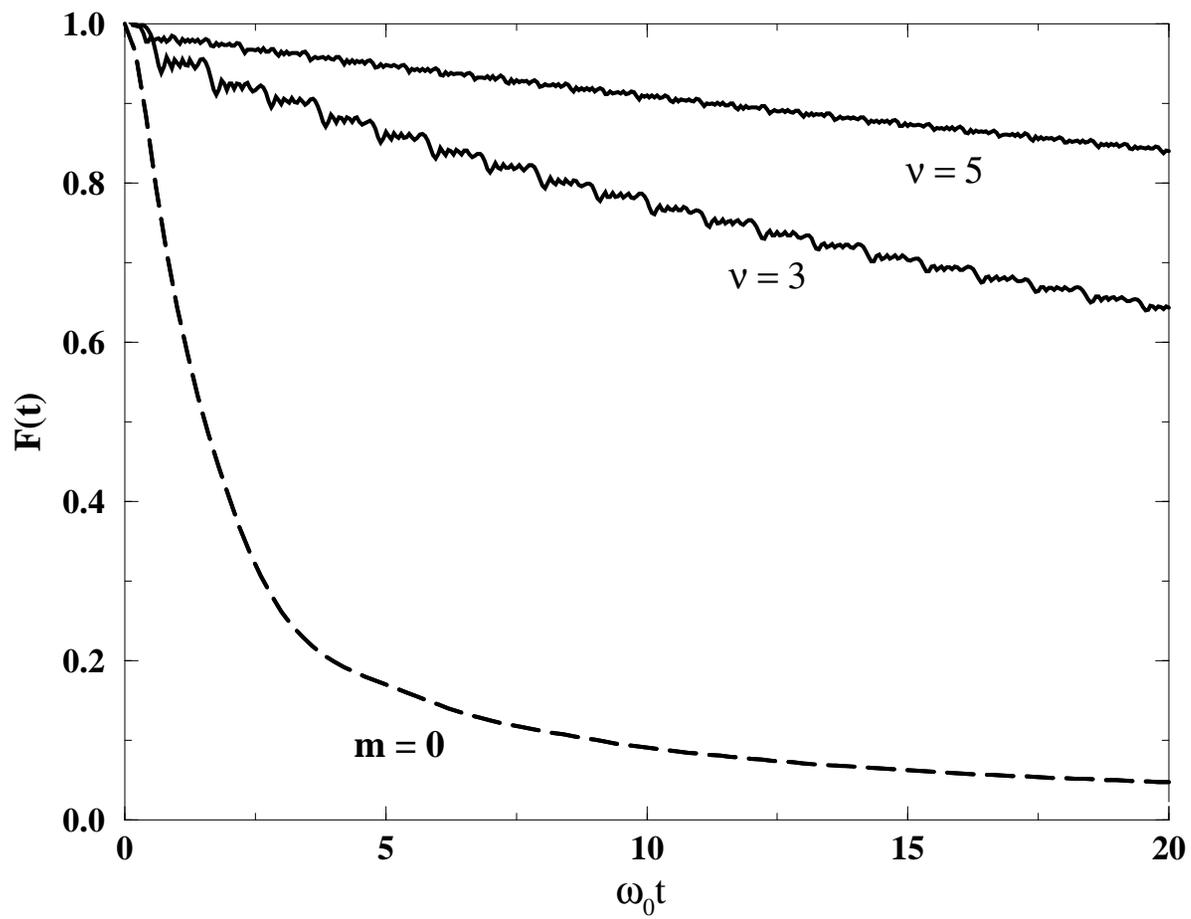
\end{document}